\documentclass[traditabstract]{aa} 

\usepackage[colorlinks=true, linkcolor=blue, citecolor=blue, urlcolor=blue]{hyperref}
\usepackage{amsmath}
\usepackage{graphicx}
\usepackage{txfonts}
\usepackage{natbib}
\usepackage{upgreek}

\begin{document}

   \title{Multi-wavelength study of the pre-eruption dip in the recurrent nova T Coronae Borealis preceding imminent nova eruption}
 \author{Songpeng Pei\inst{1}
          \and
          Xiaowan Zhang\inst{1}
          \and
          Renzhi Su\inst{2}
          \and
          Yongzhi Cai\inst{3, 4, 5}
          \and
          Ziwei Ou\inst{6}
          \and
          Qiang Li\inst{7, 8}
          \and
          Xiaoqin Ren\inst{7, 8}
          \and
          Taozhi Yang\inst{9}
          \and
          Mingyue Li\inst{7, 8}
          }

   \institute{School of Physics and Electrical Engineering, Liupanshui Normal University, Liupanshui, Guizhou, 553004, China\\
         \email{songpengpei@outlook.com}
         \and
             Shanghai Astronomical Observatory, Chinese Academy of Sciences, 80 Nandan Road, Shanghai 200030, China
         \and
             Yunnan Observatories, Chinese Academy of Sciences, Kunming 650216, China
         \and
             International Centre of Supernovae, Yunnan Key Laboratory, Kunming 650216, China
         \and
             Key Laboratory for the Structure and Evolution of Celestial Objects, Chinese Academy of Sciences, Kunming 650216, China
          \and
             Tsung-Dao Lee Institute, Shanghai Jiao Tong University, Shanghai 201210, China
         \and
             School of Physics and Electronic Science, Qiannan Normal University for Nationalities, Duyun 558000, China    
         \and
             Qiannan Key Laboratory of Radio Astronomy, Guizhou Province, Duyun 558000, China
         \and
             Ministry of Education Key Laboratory for Nonequilibrium Synthesis and Modulation of Condensed Matter, School of Physics, Xi'an Jiaotong University, 710049 Xi'an, China           
             }
   \date{Received 00.00.2025; accepted 00.00.2025}
   
\abstract{We present a multi-wavelength study of the symbiotic recurrent nova (RN) T Coronae Borealis (T CrB) using Swift Burst Alert Telescope (BAT)/ X-Ray Telescope (XRT) / UltraViolet Optical Telescope (UVOT) and American Association of Variable Stars Observers (AAVSO) observations from 2005 to 2025. Our analysis spans quiescent, high, and pre-eruption dip states. We find that brightening amplitudes increase toward shorter wavelengths in both optical and UV bands, while the UV and X-ray fluxes are generally anti-correlated throughout all phases. During the 2023-2024 pre-eruption dip, soft and hard X-rays increased as optical and ultraviolet (UV) brightness declined, consistent with a transition from an optically thick to thin boundary layer driven by a reduction in the accretion rate. We also report, for the first time, a second, lower-amplitude dip occurring between September 2024 and February 2025 following the primary 2023–2024 pre-eruption dip. The observed variability supports an accretion-variation scenario as a unifying explanation for both the high and dip states, and may signal an imminent nova eruption.}

\keywords{accretion, accretion disks --- stars: binaries: symbiotic --- stars: novae, cataclysmic variables --- stars: individual (T CrB) --- stars: white dwarfs--- X-rays: binaries}

%
\titlerunning{Pre-eruption dip in T Coronae Borealis}
\maketitle

\section{Introduction} \label{sec:intro}
T Coronae Borealis (T CrB; HR 5958; HD 143454) is the nearest known RN and one of the brightest and most studied in this class. It has had recorded eruptions in 1866 and 1946, each peaking at about $V\approx2.0$ \citep{1949ApJ...109...81S}. There are also possible historical eruptions in AD 1217 and 1787 \citep{2023JHA....54..436S, 2023ATel16107....1S}. After $\sim$ 80 years of quiescence since 1946, the next nova eruption is widely predicted to occur imminently \citep{2016NewA...47....7M, 2023ApJ...953L...7I, 2023AstL...49..501M, 2023A&A...680L..18Z, 2025MNRAS.541L..14M}, and current estimates place it in the 2023.0–2026.8 range \citep{2020ApJ...902L..14L, 2023ATel16107....1S, 2023MNRAS.524.3146S}. In general, recurrent novae (RNe) are rare (only $\sim$ 11 are known in the Milky Way), and T CrB is the nearest of the four known symbiotic RNe. This imminent outburst could become the brightest nova of the modern era.

T CrB comprises a Roche lobe–filling M4 III red giant ($M = 0.69^{+0.02}_{-0.01} M_{\odot}$, $R = 65\pm6 R_\odot$) \citep{1998MNRAS.296...77B, 1999A&AS..137..473M, 2025ApJ...983...76H} and a extremely massive white dwarf (WD; M$_{WD}$ = 1.37 $\pm$ 0.01~$M_\odot$; \citep{2025ApJ...983...76H}). The orbital period is 227.57-227.58 d \citep{2000AJ....119.1375F, 2023MNRAS.524.3146S, 2025A&A...694A..85P}, and the $V$-band light curve shows a 0.3 mag ellipsoidal modulation at half the orbital period \citep{1975JBAA...85..217B, 2023MNRAS.524.3146S}. The distance is estimated as $896\pm22$ pc\footnote{http://dc.g-vo.org/tableinfo/gedr3dist.main} from {\it Gaia} early data release 3 (eDR3) \citep{2021AJ....161..147B}. \citet{2022MNRAS.517.6150S} estimated a similar distance of $\sim914$ pc. T CrB has also been suggested to display dwarf-nova–like outbursts under some conditions \citep{2023ApJ...953L...7I}.

Since its 1946 nova event, T CrB has alternated between long quiescent intervals and episodes of enhanced activity. Around 2014–2015 the system suddenly brightened: in 2015 it entered a so-called super-active state \footnote{Throughout this text, the term {\it high} state refers to the period of heightened activity in T CrB that began in 2014–2015 and ended in 2023} akin to that seen in the few years before the 1946 eruption \citep{2016NewA...47....7M, 2016MNRAS.462.2695I, 2018A&A...619A..61L}. During this high state the star’s UV, optical, and radio brightness increased dramatically; the usual orbital modulation in the $B$ band disappeared; strong high-ionization emission lines appeared in the spectrum; and the hard X-ray flux all but vanished \citep{2016NewA...47....7M, 2018A&A...619A..61L, 2019ApJ...884....8L}. It has been suggested that this high state may be driven by a SU UMa–type dwarf-nova mechanism \citep{2023ApJ...953L...7I}. This unusual behavior has been suggested as a possible precursor of the next nova eruption \citep{2016NewA...47....7M, 2020ApJ...902L..14L, 2023MNRAS.524.3146S, 2023ApJ...953L...7I, 2023A&A...680L..18Z}. 

By mid-2023 the high state was clearly winding down \citep{2023RNAAS...7..145M, 2023ATel16109....1T}: the optical luminosity fell from about 20 $-$ 25 $L_\odot$ in April–May 2023 to $\sim$ 8 $-$ 9 $L_\odot$ by August \citep{2023A&A...680L..18Z}. In April 2023 the system entered a pre-eruption dip state \footnote{Throughout this text, the term {\it pre-eruption dip} refers to the phase of T CrB's activity that began in 2023 and ended in 2024} \citep{2023ATel16107....1S, 2023ATel16114....1K}, as had been predicted by \citet{2023MNRAS.524.3146S}. In this dip, the optical and UV flux declined noticeably – a behavior reminiscent of the fading seen before the 1946 nova eruption. Like the high state, the 2023 dip was proposed as a potential precursor to the forthcoming nova eruption \citep{2023MNRAS.524.3146S, 2023ATel16107....1S, 2023BAAVC.196....8T, 2023ATel16114....1K}.

However, \citet{2025MNRAS.541L..14M} have shown that the 2023–2024 pre-eruption dip only partially resembled the 1946 event. In particular, the dip matched the 1946 behavior in the $B$ band but not in the $V$ band, leading them to argue that it may not reliably predict the nova timing. These deviations imply that the next nova eruption could occur somewhat earlier or later than the date predicted based on the pre-eruption dip \citep{2023ATel16107....1S}. In any case, the 2023–2024 pre-eruption dip – observed in detail with modern multiwavelength instrumentation – offers valuable insight into the accretion physics of this symbiotic RN.

T CrB thus occupies a unique niche among interacting binaries. It shows features of RNe, symbiotic stars, cataclysmic variables, and dwarf novae all at once. Its near-Chandrasekhar-limit WD, together with phenomena such as the high state, the pre-eruption dip, and the post-nova secondary maximum, make it an important laboratory for studying accretion in long-period WD binaries. In particular, constraining its behavior helps inform models of binary evolution leading to Type Ia supernovae \citep{1997MNRAS.288.1027S, 2000AJ....119.1375F, 2001ApJ...558..323H, 2019A&A...622A..35L}.

The high state observed between 2014 and 2023 has been extensively studied \citep[e.g.,][]{2016NewA...47....7M, 2016MNRAS.462.2695I, 2019MNRAS.489.2930Z, 2019ApJ...884....8L, 2020ApJ...902L..14L, 2023A&A...680L..18Z, 2023RNAAS...7..145M, 2023MNRAS.524.3146S, 2023ApJ...953L...7I, 2025BlgAJ..42...29S, 2025MNRAS.541L..14M, 2025A&A...694A..85P, 2025ApJ...989...78S}. In this paper we analyze archival X-ray, optical, and UV observations of T CrB from 2005–2025 (spanning the quiescent, high, and pre-dip states) in order to investigate the nature of the pre-eruption dip. Our goal is to use this multiwavelength dataset to shed light on the accretion processes and impending nova behavior of this iconic symbiotic RN.

The UVW1, UVM2, and UVW2 filter data from Swift/UVOT used in this study were previously analyzed by \citet{2025A&A...701A.176M}. Portions of the Swift/XRT dataset have been examined by \citet{2009ApJ...701.1992K, 2016MNRAS.462.2695I, 2018A&A...619A..61L, 2019MNRAS.489.2930Z}, while segments of the Swift/BAT observations were analyzed by \citet{2018A&A...619A..61L}. In this paper, we build on these works by presenting new findings on the iconic and extensively studied RN T CrB. In Sect. \ref{sec:intro}, we provide a brief introduction to T CrB. In Section~\ref{sec:observation}, we review the details of the observations and the data reduction process. In Section~\ref{sec:timing}, we present results from analysis of the X-ray, optical, and UV emission of T CrB. In Section~\ref{sec:discussion}, we discuss our findings and Section~\ref{sec:conclusions} presents our summary and conclusions.

\section{Observation and data reduction} \label{sec:observation}
The Neil Gehrels Swift Observatory \citep[hereafter, {\em Swift};][]{2004ApJ...611.1005G} began observing T CrB on 2005-06-17. T CrB was monitored with the Swift using all three instruments: the XRT \citep{2005SSRv..120..165B}, the BAT \citep{2005SSRv..120..143B}, and the UVOT \citep{2005SSRv..120...95R}. We used Swift observations from 2005-06-17 to 2025-05-03 (MJD 53538–60798), comprising 278 pointed observations (total XRT exposure: 437.3 ks), as shown in Table~\ref{tab:obsT CrB}. The XRT monitoring operated exclusively in Photon Counting (PC) mode. Raw XRT event files were processed with FTOOLS package in HEASOFT v. 6.30.1\footnote{https://heasarc.gsfc.nasa.gov/docs/software/lheasoft/download.html} using the xrtpipeline task and the latest calibration files. Source events were then extracted with XSELECT from a circular region of radius 20 pixels centered on the SIMBAD coordinates of T CrB, and background events were taken from a nearby source-free region. The resulting XRT count rates were corrected for vignetting and bad CCD columns using the XRTLCCORR tool. The processed BAT data obtained from the Swift BAT transient monitor page was also used \citep{2013ApJS..209...14K}. These BAT data extend the hard X-ray coverage and trace the long-term high-energy behavior of the system.

Simultaneously, each Swift pointing included UVOT exposures. During 99 observations the Swift UVOT \citep{2005SSRv..120...95R} was in the event mode, providing fast photometry in the UVM2 filter (90 observations) and UGRISM filter (9 observations). Swift UVOT was in the image mode in the rest 179 observations, providing the mean magnitude of each observation in one or more of the six UVOT filters (V, B, U, UVW1, UVM2 and UVW2). All the Swift UVOT data were processed by using the FTOOLS package. UVOT data were reduced using the standard HEASoft/UVOT pipeline. Source brightness and count rates were extracted with uvotsource using a 5$''$-radius aperture, with the background estimated from a nearby sky region. Some UVOT frames were saturated (source brighter than around 10–11 mag), so for those exposures we measured the brightness using the read-out streak method \citep{2013MNRAS.436.1684P}. Because the Swift/UVOT UVW2 filter is affected by a significant red leak, and T CrB has a bright, very red component from its red giant companion, the leaked optical light dominates the UV signal—especially during quiescent and pre-eruption dip states. Consequently, the UVW2 light curve presented here is included for reference only.

The logs of all the observations with the date, exposure time and mean count rate are compiled in Table~\ref{tab:obsT CrB}. 

The high-energy observations were supplemented with long-term optical photometry from the AAVSO. We retrieved all publicly available I, R, V, B and U bands observations of T CrB from the AAVSO International Database, covering roughly 2005–2025 ($\sim$ 7400 days).

All timing data (X-ray) were referred to the Solar System barycenter (SSB) using the FTOOLS barycorr task. The multi-instrument data set (Swift XRT/BAT/UVOT plus AAVSO photometry) thus provides a well-calibrated, contemporaneous record of T CrB activity prior to its expected nova outburst.

\setcounter{table}{0}
\begin{table}
\begin{minipage}{80mm}
\caption{Log of all Swift XRT (PC mode) observations of T CrB between 2005.06 and 2025.05.}
\label{tab:obsT CrB}
\begin{tabular}{lccc}
\hline
\hline

      ObsID & Date$^{a}$ & Exp. (s)& Count rate  \\
\hline
00035171001 &53538.42  &8822.9  &0.075 $\pm$ 0.009  \\
00035171002 &53650.70  &5185.6  &0.080 $\pm$ 0.006  \\
00035171003 &53662.33  &10325.2  &0.101 $\pm$ 0.004  \\
00035171004 &54538.42  &4033.8  &0.109  $\pm$ 0.007  \\
00035171005 &54540.06  &4876.8  &0.099  $\pm$ 0.005  \\
00035171006 &54541.99  &4080.5  &0.101 $\pm$ 0.006  \\
00045776001 &55892.82  &2730.3  &0.063 $\pm$ 0.005  \\
00045776002 &55899.40  &1236.1  &0.101 $\pm$ 0.010  \\
00081659001 &57288.40  &8255.7  &0.063 $\pm$ 0.003  \\
00081659002 &57289.12  &1657.3  &0.071 $\pm$ 0.008  \\
00045776003 &57296.06  &193.0  &0.023 $\pm$ 0.005  \\
00045776005 &57771.52  &10023.8  &0.012 $\pm$ 0.001  \\
00045776006 &57783.37  &9390.9  &0.016 $\pm$ 0.002  \\
00045776007 &57792.74  &475.9  &0.019 $\pm$ 0.009  \\
00045776008 &57793.51  &3269.8  &0.018 $\pm$ 0.003  \\
00045776009 &57794.91  &1313.8  &0.023 $\pm$ 0.005  \\
00045776010 &57797.16  &1125.9  &0.026 $\pm$ 0.006  \\
00045776011 &57802.58  &7092.9  &0.015 $\pm$ 0.002  \\
00045776012 &57806.93  &987.9  &0.011 $\pm$ 0.004  \\
00045776014 &57815.63  &498.7  &0.010 $\pm$ 0.007  \\
00045776015 &57820.90  &2810.7  &0.014 $\pm$ 0.003  \\
00045776016 &57824.82  &5390.3  &0.015 $\pm$ 0.002  \\
00045776017 &57834.52  &2663.1  &0.012 $\pm$ 0.003  \\
00045776018 &57850.68  &3882.3  &0.007 $\pm$ 0.002  \\
00045776019 &57863.12  &238.2  &0.017 $\pm$ 0.012  \\
00045776020 &57864.93  &4021.0  &0.011 $\pm$ 0.002  \\
00045776021 &57885.48  &3791.8  &0.006 $\pm$ 0.002  \\
00045776022 &57932.74  &3557.9  &0.012 $\pm$ 0.002  \\
00045776023 &57952.68  &3919.9  &0.007 $\pm$ 0.002  \\
00045776024 &57987.44  &1458.5  &0.016 $\pm$ 0.004  \\
00045776025 &57988.91  &1524.4  &0.019 $\pm$ 0.005  \\
00045776026 &58022.53  &4022.3  &0.019 $\pm$ 0.003  \\
00045776027 &58053.45  &743.3  &0.005 $\pm$ 0.004  \\
00045776028 &58055.25  &1697.4  &0.009 $\pm$ 0.003  \\
00045776029 &58122.07  &3640.3  &0.016 $\pm$ 0.002  \\
00045776030 &58174.40  &4068.4  &0.004 $\pm$ 0.001  \\
00045776031 &58257.56  &5067.3  &0.009 $\pm$ 0.002  \\
00045776032 &58326.26  &556.7  &0.008 $\pm$ 0.002  \\
00045776036 &58391.61  &461.3  &0.013 $\pm$ 0.002  \\
00045776038 &58471.73  &807.3  &0.007 $\pm$ 0.001  \\
00045776040 &58548.50  &656.9  &0.010 $\pm$ 0.002  \\
00011405001 &58623.67  &366.7  &0.008 $\pm$ 0.002  \\
00045776042 &58668.57  &268.2  &0.025 $\pm$ 0.005  \\
00045776044 &58671.89  &213.0  &0.021 $\pm$ 0.005  \\
00011548001 &58730.38  &366.0  &0.010 $\pm$ 0.002  \\
00012011001 &58754.55  &396.1  &0.005 $\pm$ 0.002  \\
00013032002 &58839.51  &1459.2  &0.016 $\pm$ 0.004  \\
00013032003 &58842.04  &280.8  &0.013 $\pm$ 0.003  \\
00013294047 &58929.42  &398.6  &0.013 $\pm$ 0.003  \\
00013559001 &59011.34  &353.5  &0.019 $\pm$ 0.003  \\
00013558002 &59044.37  &396.5  &0.023 $\pm$ 0.003  \\
00013558004 &59072.22  &446.3  &0.022 $\pm$ 0.004  \\
00089009001 &59100.70  &2808.9  &0.011 $\pm$ 0.002  \\
00013922001 &59189.04  &195.6  &0.007 $\pm$ 0.003  \\
00014166005 &59287.69  &142.9  &0.038 $\pm$ 0.007  \\
00014166007 &59289.94  &183.0  &0.030 $\pm$ 0.006  \\
00014166009 &59290.67  &223.1  &0.020 $\pm$ 0.005  \\
00014379001 &59381.47  &290.8  &0.034 $\pm$ 0.006  \\
00012011005 &59405.37  &150.4  &0.011 $\pm$ 0.005  \\
00012011007 &59411.57  &411.2  &0.032 $\pm$ 0.005  \\
00012011009 &59439.16  &160.5  &0.010 $\pm$ 0.007  \\
00012011011 &59442.34  &178.0  &0.011 $\pm$ 0.004  \\
00014765003 &59443.53  &153.0  &0.011 $\pm$ 0.004  \\
00014765005 &59447.06  &225.6  &0.032 $\pm$ 0.006  \\
\hline
\multicolumn{4}{p{.9\textwidth}}{\textbf{Notes: }$^a$ Modified Julian Date. The count rates were measured in the 0.3--10.0 keV.}\\

\end{tabular}
\end{minipage} 
\end{table}

%
\setcounter{table}{0}
\begin{table}
\begin{minipage}{80mm}
\caption{Continued. Log of all Swift XRT (PC mode) observations of T CrB between 2005.06 and 2025.05.}
\label{tab:obsT CrB}
\begin{tabular}{lccc}
\hline
\hline

      ObsID & Date$^{a}$ & Exp. (s)& Count rate  \\
\hline
00012011013 &59450.11  &203.2  &0.031 $\pm$ 0.008  \\
00013558010 &59646.60  &152.9  &0.020 $\pm$ 0.007  \\
00013558012 &59646.99  &1053.0  &0.012 $\pm$ 0.004  \\
00013558013 &59649.06  &223.0  &0.016 $\pm$ 0.005  \\
00013558016 &59721.49  &581.4  &0.029 $\pm$ 0.011  \\
00013558017 &59752.33  &198.1  &0.010 $\pm$ 0.005  \\
00013558021 &59782.49  &195.6  &0.029 $\pm$ 0.006  \\
00013558023 &59790.12  &193.1  &0.038 $\pm$ 0.008  \\
00013558026 &59791.31  &163.0  &0.027 $\pm$ 0.006  \\
00013558028 &59793.50  &225.7  &0.005 $\pm$ 0.003  \\
00013558030 &59798.48  &178.0  &0.022 $\pm$ 0.007  \\
00013558032 &59923.21  &82.2  &0.019 $\pm$ 0.005  \\
00013558034 &59923.31  &1281.2  &0.023 $\pm$ 0.005  \\
00013558036 &59923.68  &218.1  &0.017 $\pm$ 0.005  \\
00013558035 &59923.81  &792.4  &0.024 $\pm$ 0.006  \\
00013558040 &59954.09  &1043.0  &0.030 $\pm$ 0.006  \\
00013558038 &59954.15  &188.2  &0.043 $\pm$ 0.009  \\
00013558042 &59954.69  &165.5  &0.020 $\pm$ 0.005  \\
00013558041 &59954.75  &712.1  &0.016 $\pm$ 0.006  \\
00013558046 &59985.58  &1211.1  &0.051 $\pm$ 0.005  \\
00013558044 &59986.34  &233.2  &0.027 $\pm$ 0.006  \\
00013558053 &60013.68  &744.7  &0.035 $\pm$ 0.009  \\
00013558052 &60014.68  &1023.0  &0.070 $\pm$ 0.014  \\
00013558056 &60044.32  &1547.0  &0.044 $\pm$ 0.005  \\
00013558059 &60044.58  &937.7  &0.073 $\pm$ 0.013  \\
00013558057 &60044.66  &682.0  &0.092 $\pm$ 0.013  \\
00013558060 &60074.30  &218.1  &0.033 $\pm$ 0.011  \\
00013558065 &60074.89  &403.7  &0.018 $\pm$ 0.009  \\
00013558067 &60076.79  &461.3  &0.042 $\pm$ 0.015  \\
00013558068 &60080.50  &1459.3  &0.016 $\pm$ 0.004  \\
00013558069 &60089.75  &105.3  &0.013 $\pm$ 0.010  \\
00013558072 &60089.97  &747.2  &0.004 $\pm$ 0.007  \\
00013558077 &60104.45  &1353.8  &0.025 $\pm$ 0.005  \\
00013558083 &60119.33  &1441.7  &0.023 $\pm$ 0.005  \\
00013558081 &60119.45  &170.5  &0.019 $\pm$ 0.009  \\
00013558087 &60119.72  &275.8  &0.023 $\pm$ 0.008  \\
00013558084 &60119.78  &646.9  &0.031 $\pm$ 0.009  \\
00013558091 &60134.09  &1373.5  &0.013 $\pm$ 0.004  \\
00013558089 &60134.19  &135.4  &0.025 $\pm$ 0.007  \\
00013558093 &60134.73  &228.2  &0.012 $\pm$ 0.005  \\
00013558092 &60134.98  &491.5  &0.030 $\pm$ 0.010  \\
00013558095 &60149.25  &130.4  &0.004 $\pm$ 0.004  \\
00013558098 &60149.59  &777.3  &0.002 $\pm$ 0.003  \\
00013558101 &60149.66  &195.6  &0.010 $\pm$ 0.004  \\
00013558103 &60150.52  &1338.0  &0.012 $\pm$ 0.004  \\
00013558106 &60164.40  &1502.2  &0.107 $\pm$ 0.006  \\
00013558110 &60178.44  &1562.0  &0.087 $\pm$ 0.007  \\
00013558116 &60178.88  &208.1  &0.043 $\pm$ 0.008  \\
00013558118 &60191.55  &1073.1  &0.124 $\pm$ 0.012  \\
00013558121 &60192.08  &504.0  &0.153 $\pm$ 0.019  \\
00013558124 &60192.15  &180.5  &0.130 $\pm$ 0.017  \\
00013558126 &60192.54  &170.5  &0.124 $\pm$ 0.012  \\
00013558128 &60206.16  &1393.6  &0.119 $\pm$ 0.010  \\
00013558131 &60206.42  &1451.7  &0.122 $\pm$ 0.013  \\
00013558132 &60209.06  &62.7  &0.109 $\pm$ 0.016  \\
00013558136 &60220.49  &1537.0  &0.147 $\pm$ 0.007  \\
00013558148 &60234.43  &152.9  &0.130 $\pm$ 0.016  \\
00013558147 &60234.62  &586.7  &0.099 $\pm$ 0.015  \\    
00013558144 &60234.81  &1075.6  &0.095 $\pm$ 0.011  \\
00013558150 &60234.88  &140.4  &0.082 $\pm$ 0.010  \\
00013558153 &60245.05  &268.3  &0.084 $\pm$ 0.006  \\ 
00013558152 &60245.48  &817.5  &0.053 $\pm$ 0.004  \\
00013558158 &60276.21  &198.1  &0.064 $\pm$ 0.008  \\
00013558160 &60276.63  &1737.1  &0.105 $\pm$ 0.006  \\
\hline
\multicolumn{4}{p{.9\textwidth}}{{\bf Notes: }$^a$ Modified Julian Date. The count rates were measured in the 0.3 $-$ 10.0 keV.}\\

\end{tabular}
\end{minipage} 
\end{table}

%
\setcounter{table}{0}
\begin{table}
\begin{minipage}{80mm}
\caption{Continued. Log of all Swift XRT (PC mode) observations of T CrB between 2005.06 and 2025.05.}
\label{tab:obsT CrB}
\begin{tabular}{lccc}
\hline
\hline

      ObsID & Date$^{a}$ & Exp. (s)& Count rate  \\
\hline
00013558164 &60290.26  &185.2  &0.107 $\pm$ 0.015  \\
00013558168 &60290.84  &183.2  &0.079 $\pm$ 0.009  \\
00013558167 &60290.92  &819.9  &0.045 $\pm$ 0.009  \\
00013558170 &60304.03  &255.7  &0.169 $\pm$ 0.017  \\
00013558174 &60304.10  &250.7  &0.099 $\pm$ 0.017  \\
00013558173 &60304.82  &661.9  &0.058 $\pm$ 0.010  \\
00013558176 &60318.35  &208.1  &0.055 $\pm$ 0.013  \\ 
00013558180 &60318.76  &195.6  &0.100 $\pm$ 0.010  \\
00013558185 &60332.07  &812.4  &0.106 $\pm$ 0.013  \\
00013558189 &60332.33  &892.6  &0.093 $\pm$ 0.012  \\
00013558182 &60332.53  &1328.9  &0.053 $\pm$ 0.009  \\
00013558188 &60332.92  &917.7  &0.044 $\pm$ 0.008  \\
00013558191 &60346.11  &205.6  &0.054 $\pm$ 0.012  \\
00013558193 &60346.78  &498.9  &0.092 $\pm$ 0.016  \\
00013558194 &60346.84  &200.6  &0.031 $\pm$ 0.011 \\
00013558202 &60362.14  &438.8  &0.069 $\pm$ 0.017  \\
00013558200 &60362.78  &152.9  &0.091 $\pm$ 0.014  \\
00013558204 &60376.36  &183.0  &0.068 $\pm$ 0.011  \\
00013558206 &60376.82  &1381.5  &0.090 $\pm$ 0.007  \\
00097564001 &60411.01  &92.8  &0.085 $\pm$ 0.016  \\
00097564005 &60411.73  &178.0  &0.051 $\pm$ 0.011  \\
00097564004 &60411.86  &897.6  &0.051 $\pm$ 0.009  \\
00097564009 &60421.53  &769.7  &0.032 $\pm$ 0.008  \\
00097564014 &60421.87  &100.3  &0.029 $\pm$ 0.009  \\
00097564016 &60431.21  &125.4  &0.017 $\pm$ 0.005  \\
00097564015 &60431.42  &939.9  &0.067 $\pm$ 0.010  \\
00097564018 &60431.74  &183.0  &0.036 $\pm$ 0.007  \\
00097564021 &60441.41  &160.5  &0.047 $\pm$ 0.009  \\
00097564020 &60441.49  &860.0  &0.103 $\pm$ 0.013  \\
00097564023 &60441.62  &897.6  &0.029 $\pm$ 0.012 \\
00097564024 &60441.81  &162.8  &0.042 $\pm$ 0.015  \\
00097564027 &60451.48  &213.1  &0.078 $\pm$ 0.009  \\
00097564029 &60451.81  &123.0  &0.068 $\pm$ 0.008  \\
00097564026 &60451.87  &1501.9  &0.110 $\pm$ 0.011  \\
00097564033 &60461.35  &190.5  &0.102 $\pm$ 0.011  \\
00097564031 &60461.49  &952.8  &0.122 $\pm$ 0.013  \\
00097564035 &60461.74  &190.5  &0.108 $\pm$ 0.027  \\
00097564036 &60461.75  &957.8  &0.099 $\pm$ 0.019  \\
00097564037 &60471.28  &195.6  &0.094 $\pm$ 0.011  \\
00097564041 &60471.54  &163.0  &0.053 $\pm$ 0.008  \\
00097564040 &60471.67  &718.8  &0.074 $\pm$ 0.012  \\
00097564039 &60471.80  &890.1  &0.081 $\pm$ 0.011  \\
00097564043 &60481.34  &144.1  &0.060 $\pm$ 0.009  \\
00097564045 &60481.40  &865.0  &0.048 $\pm$ 0.009  \\
00097564047 &60481.55  &195.6  &0.058 $\pm$ 0.009  \\
00097564046 &60481.73  &265.8  &0.014 $\pm$ 0.011  \\
00097564049 &60491.33  &165.5  &0.114 $\pm$ 0.012  \\
00097564051 &60491.40  &967.8  &0.082 $\pm$ 0.011  \\
00097564053 &60491.66  &165.5  &0.070 $\pm$ 0.009  \\
00097564052 &60491.86  &927.7  &0.034 $\pm$ 0.007  \\
00097564058 &60501.40  &581.7  &0.027 $\pm$ 0.015  \\
00097564057 &60501.46  &863.8  &0.051 $\pm$ 0.016  \\
00097564059 &60501.72  &120.3  &0.046 $\pm$ 0.008  \\
00097564064 &60511.31  &743.6  &0.021 $\pm$ 0.008  \\
00097564061 &60511.38  &205.6  &0.049 $\pm$ 0.009  \\  
00097564065 &60511.65  &117.8  &0.063 $\pm$ 0.012  \\
00097564063 &60511.78  &885.1  &0.045 $\pm$ 0.010  \\
00097564069 &60521.31  &882.6  &0.024 $\pm$ 0.006  \\
00097564067 &60521.38  &160.6  &0.051 $\pm$ 0.010  \\
00097564070 &60521.44  &819.9  &0.044 $\pm$ 0.008  \\ 
00097564071 &60521.97  &240.7  &0.055 $\pm$ 0.010  \\
00097564073 &60531.02  &117.8  &0.014 $\pm$ 0.006  \\
00097564076 &60531.35  &423.0  &0.018 $\pm$ 0.009  \\
00097564077 &60531.68  &195.6  &0.019 $\pm$ 0.006  \\
\hline
\multicolumn{4}{p{.9\textwidth}}{{\bf Notes: }$^a$ Modified Julian Date. The count rates were measured in the 0.3 -- 10.0 keV.}\\

\end{tabular}
\end{minipage} 
\end{table}

%
\setcounter{table}{0}
\begin{table}
\begin{minipage}{80mm}
\caption{Continued. Log of all Swift XRT (PC mode) observations of T CrB between 2005.06 and 2025.05.}
\label{tab:obsT CrB}
\begin{tabular}{lccc}
\hline
\hline

      ObsID & Date$^{a}$ & Exp. (s)& Count rate  \\
\hline
00097564075 &60531.82  &920.2  &0.013 $\pm$ 0.005  \\
00097564082 &60541.20  &60.2  &0.014 $\pm$ 0.005  \\
00097564084 &60541.27  &769.7  &0.043 $\pm$ 0.009  \\
00097564085 &60541.33  &777.3  &0.012 $\pm$ 0.005  \\
00097564086 &60541.60  &55.2  &0.019 $\pm$ 0.010  \\
00089722001 &60546.79  &1912.0  &0.025 $\pm$ 0.004  \\
00097564088 &60551.05  &937.8  &0.022 $\pm$ 0.006  \\
00097564093 &60551.12  &225.6  &0.033 $\pm$ 0.010  \\
00097564091 &60551.51  &208.0  &0.017 $\pm$ 0.005  \\
00097564089 &60551.59  &910.1  &0.024 $\pm$ 0.006  \\
00097564094 &60561.29  &72.7   &0.077 $\pm$ 0.044  \\
00097564096 &60561.44  &303.4  &0.033 $\pm$ 0.021  \\
00097564098 &60561.70  &155.4  &0.021 $\pm$ 0.008  \\
00097564111 &60571.07  &72.7   &0.014 $\pm$ 0.006  \\
00097564107 &60571.46  &621.8  &0.030 $\pm$ 0.008  \\
00097564105 &60571.53  &75.2   &0.019 $\pm$ 0.006  \\
00097564109 &60571.85  &824.9  &0.023 $\pm$ 0.006  \\
00097564112 &60581.24  &168.0  &0.046 $\pm$ 0.008  \\
00097564114 &60581.38  &930.2  &0.019 $\pm$ 0.006  \\
00097564115 &60581.44  &877.4  &0.035 $\pm$ 0.007  \\
00097564116 &60581.70  &135.4  &0.022 $\pm$ 0.007  \\
00097564120 &60591.06  &1100.7 &0.026 $\pm$ 0.006  \\
00097564118 &60591.19  &142.9  &0.018 $\pm$ 0.010  \\
00097564122 &60591.65  &102.8  &0.015 $\pm$ 0.011  \\
00097564121 &60591.72  &952.8  &0.011 $\pm$ 0.005  \\
00097564124 &60601.03  &188.2  &0.009 $\pm$ 0.006  \\
00097564126 &60601.56  &1073.1 &0.021 $\pm$ 0.005  \\
00097564128 &60601.95  &198.0  &0.017 $\pm$ 0.008  \\
00097564132 &60611.05  &887.6  &0.015 $\pm$ 0.005  \\
00097564133 &60611.38  &702.0  &0.027 $\pm$ 0.007  \\
00097564130 &60611.45  &73.6   &0.029 $\pm$ 0.010  \\
00097564134 &60611.59  &40.3   &0.025 $\pm$ 0.009  \\
00097564138 &60643.19  &170.5  &0.022 $\pm$ 0.008  \\
00097564136 &60643.51  &895.1  &0.018 $\pm$ 0.006  \\
00097564140 &60643.91  &135.4  &0.013 $\pm$ 0.008  \\
00097564137 &60643.98  &952.8  &0.030 $\pm$ 0.007  \\
00097564144 &60653.19  &137.9  &0.031 $\pm$ 0.011  \\
00097564143 &60653.32  &727.1  &0.031 $\pm$ 0.014  \\
00097564147 &60653.86  &293.3  &0.052 $\pm$ 0.017  \\
00097564142 &60653.91  &870.0  &0.047 $\pm$ 0.010  \\
00097564150 &60663.26  &140.4  &0.034 $\pm$ 0.013  \\
00097564152 &60663.71  &208.1  &0.025 $\pm$ 0.010  \\
00097564149 &60663.85  &920.2  &0.030 $\pm$ 0.007  \\
00097564148 &60663.98  &852.5  &0.025 $\pm$ 0.006  \\
00097564159 &60673.66  &298.0  &0.025 $\pm$ 0.012  \\
00097564155 &60673.73  &832.4  &0.017 $\pm$ 0.007  \\
00097564154 &60673.79  &865.0  &0.029 $\pm$ 0.011  \\
00097564162 &60683.12  &120.3  &0.027 $\pm$ 0.010  \\
00097564161 &60683.18  &1074.0 &0.020 $\pm$ 0.005  \\
00097564165 &60683.52  &997.9  &0.011 $\pm$ 0.009  \\
00097564160 &60683.65  &990.4  &0.015 $\pm$ 0.005  \\
00097564169 &60693.24  &210.6  &0.029 $\pm$ 0.010  \\
00097564168 &60693.38  &887.6  &0.026 $\pm$ 0.006  \\
00097564171 &60693.77  &265.8  &0.024 $\pm$ 0.009  \\
00097564167 &60693.84  &962.8  &0.016 $\pm$ 0.005  \\
00097564175 &60703.04  &218.1  &0.010 $\pm$ 0.005  \\
00097564174 &60703.10  &872.6  &0.024 $\pm$ 0.006  \\
00097564181 &60713.09  &130.4  &0.035 $\pm$ 0.013  \\
00097564183 &60713.80  &180.5  &0.050 $\pm$ 0.011  \\
00097564180 &60713.88  &614.3  &0.049 $\pm$ 0.010  \\
00097564179 &60713.94  &616.8  &0.087 $\pm$ 0.013  \\
00097564187 &60723.07  &223.1  &0.022 $\pm$ 0.007  \\
00097564186 &60723.67  &779.8  &0.031 $\pm$ 0.007  \\
00097564189 &60723.92  &283.3  &0.035 $\pm$ 0.008  \\
\hline
\multicolumn{4}{p{.9\textwidth}}{{\bf Notes: }$^a$ Modified Julian Date. The count rates were measured in the 0.3$-$10.0 keV.}\\

\end{tabular}
\end{minipage} 
\end{table}

%
\setcounter{table}{0}
\begin{table}
\begin{minipage}{80mm}
\caption{Continued. Log of all Swift XRT (PC mode) observations of T CrB between 2005.06 and 2025.05.}
\label{tab:obsT CrB}
\begin{tabular}{lccc}
\hline
\hline

      ObsID & Date$^{a}$ & Exp. (s)& Count rate  \\
\hline
00097564185 &60723.99  &918.8  &0.022 $\pm$ 0.006  \\
00097564193 &60733.44  &147.9  &0.027 $\pm$ 0.006  \\
00097564191 &60733.51  &1113.2 &0.061 $\pm$ 0.008  \\
00097564195 &60733.69  &188.1  &0.041 $\pm$ 0.008  \\
00097564192 &60733.76  &799.0  &0.064 $\pm$ 0.010  \\
00097564199 &60743.02  &27.6   &0.029 $\pm$ 0.007  \\
00097564198 &60743.16  &932.7  &0.027 $\pm$ 0.006  \\
00097564197 &60743.22  &920.2  &0.027 $\pm$ 0.006  \\
00097564205 &60753.37  &155.4  &0.034 $\pm$ 0.008  \\
00097564204 &60753.70  &877.6  &0.013 $\pm$ 0.005  \\
00097564207 &60753.77  &90.3   &0.016 $\pm$ 0.008  \\
00097564203 &60753.84  &950.3  &0.013 $\pm$ 0.005  \\
00097564212 &60763.02  &147.9  &0.035 $\pm$ 0.022  \\
00097564213 &60763.73  &93.4   &0.038 $\pm$ 0.012  \\
00097564210 &60763.87  &950.3  &0.043 $\pm$ 0.008  \\
00097564209 &60763.93  &930.2  &0.064 $\pm$ 0.010  \\
00098285001 &60768.75  &2246.7 &0.024 $\pm$ 0.004  \\
00098285005 &60783.46  &205.6  &0.022 $\pm$ 0.005  \\
00098285004 &60783.78  &2795.8 &0.048 $\pm$ 0.005  \\
00098285008 &60798.47  &190.5  &0.055 $\pm$ 0.042  \\
00098285009 &60798.47  &476.3  &0.097 $\pm$ 0.031  \\
00098285007 &60798.89  &2640.2 &0.015 $\pm$ 0.003  \\ 
\hline
\multicolumn{4}{p{.9\textwidth}}{{\bf Notes: }$^a$ Modified Julian Date. The count rates were measured in the 0.3 -- 10.0 keV.}\\

\end{tabular}
\end{minipage} 
\end{table}


\section{Multi-wavelength variability and timing analysis}
\label{sec:timing}
\subsection{Evolution of optical, UV, and X-ray light curves}
\label{sec:Multi}
\begin{figure*}\begin{center}
\includegraphics[height=15.4cm]{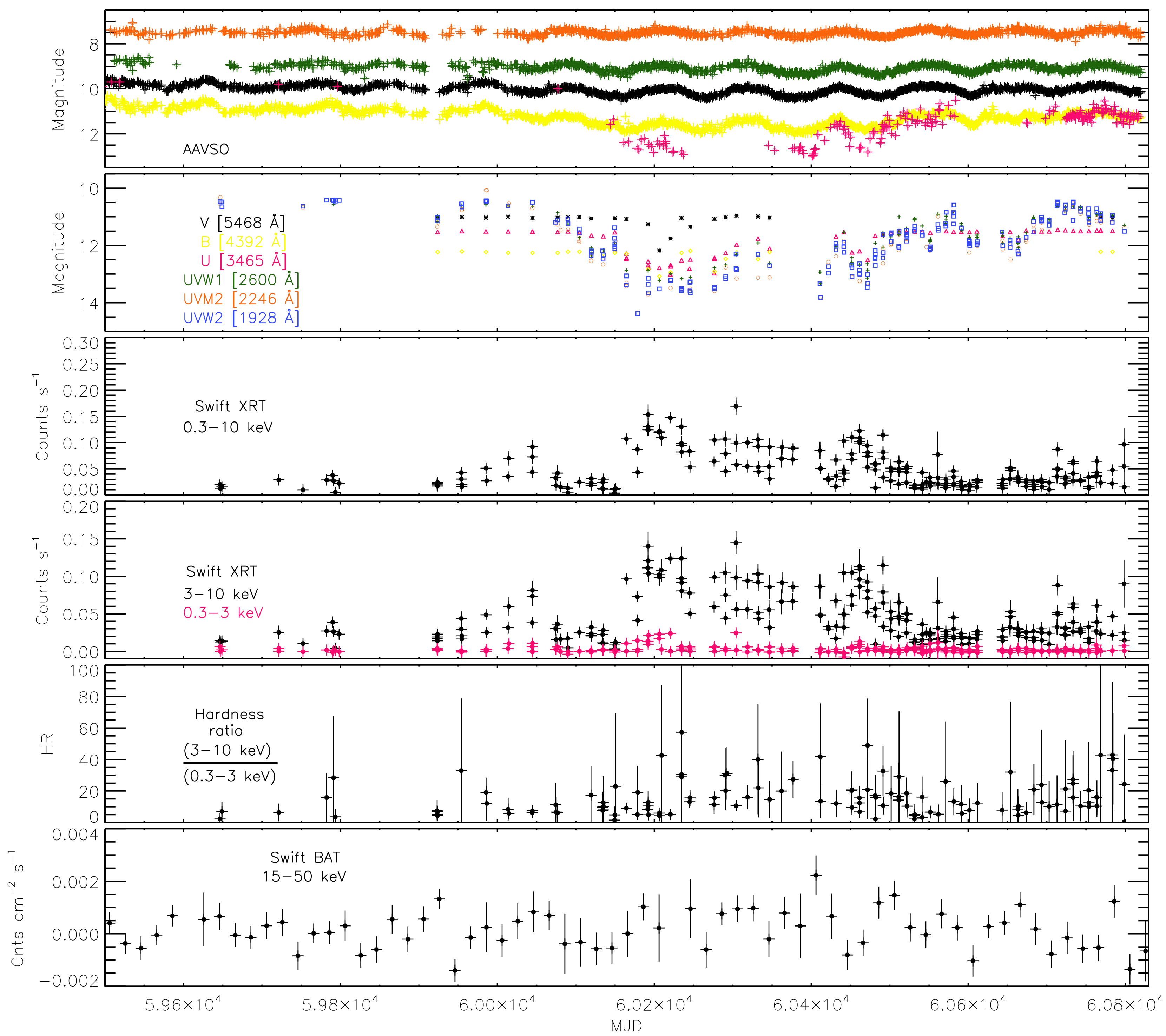}
\end{center}
\caption{From top to bottom: AAVSO light curves of T CrB from late 2021 to early 2025 in the following bands: I (green), R (orange), V (black), B (yellow), and U (red). Subsequent panels display: Swift/UVOT light curves in various filters; the Swift/XRT 0.3 -- 10 keV light curve (PC mode); Swift/XRT light curves in 3 -- 10 keV (black) and 0.3 -- 3 keV (red); the X-ray hardness ratio (3 -- 10 keV / 0.3 -- 3 keV); and the Swift/BAT light curve.}
\label{fig:sum band 02}
\end{figure*}

A comparison of T CrB’s variability across optical, UV, and X-ray energy ranges is presented in Fig.~\ref{fig:sum band 01} and Fig.~\ref{fig:sum band 02}. Fig.~\ref{fig:sum band 02} displays the same data as Fig.~\ref{fig:sum band 01}, but zoomed in from MJD 59500.0 onward to more clearly highlight the pre-eruption dip. Following \citet{2025MNRAS.541L..14M}, the AAVSO light curves have been cleaned of outliers and averaged over 1-day intervals. The Swift/UVOT UVW2 light curve is included for reference only. The Swift/BAT hard X-ray light curve has been averaged over 20-day intervals for clarity in visual presentation.

\subsubsection{Characteristics of the high state} \label{sec:high}

Following the last nova eruption, T CrB remained in a quiescent state until 2014, when it underwent a sudden optical brightening. The mass accretion rate increased abruptly by approximately a factor of 20 over quiescent levels \citep{2023MNRAS.524.3146S}, reaching $2$–$6 \times10^{-8}\rm M_\odot\rm yr^{-1}$ \citep{2018A&A...619A..61L, 2023A&A...680L..18Z, 2023MNRAS.524.3146S, 2024MNRAS.532.1421T}. This transition was accompanied by notable changes across the electromagnetic spectrum.

In X-rays, the hard component (E $\geq$ 2 keV) diminished, while a soft component (E $\leq$ 0.6 keV) emerged, indicating that the boundary layer between the WD and the inner accretion disc (AD) had become optically thick \citep{2018A&A...619A..61L, 2019ApJ...880...94L}. In the optical regime, the emergence and strengthening of high-ionization emission lines, along with a blue continuum, were observed \citep{2016NewA...47....7M}. The mean $B$-band brightness increased by 0.9 mag, deviating from the typical ellipsoidal modulation \citep{2025A&A...694A..85P}. The rising continuum corresponds to a hot component with an effective temperature of $9400 \pm 500$ K \citep{2023A&A...680L..18Z}. A simultaneous radio flux enhancement, attributed to bremsstrahlung emission from ionized circumstellar material, was also detected \citep{2019ApJ...884....8L}.

These multi-wavelength indicators confirmed the onset of a pronounced high state \citep{2016NewA...47....7M}, during which an estimated $\sim2 \times 10^{-7}~\rm M_\odot$ was accreted onto the WD \citep{2023A&A...680L..18Z}. This accumulation may be sufficient to trigger a thermonuclear runaway, with the next nova eruption predicted around 2025.5 ± 1.3 \citep{2023MNRAS.524.3146S}.

The amplitude of the brightening in the ctive state is smaller in the $V$ band than in the $B$ band \citep{2025MNRAS.541L..14M}. Moreover, at optical wavelengths, the shorter the wavelength, the greater the increase in flux compared to quiescence \citep{2016NewA...47....7M, 2016ATel.8675....1Z, 2023MNRAS.524.3146S, 2025MNRAS.541L..14M}. Thus, enhanced activity during the high state is most pronounced at the bluest optical bands.

The high state is fundamentally driven by a sharp increase in the mass transfer rate from the donor star \citep{2016MNRAS.462.2695I}. The accretion rate directly governs the optical depth of the boundary layer: higher $\dot{M}$ leads to a transition from an optically thin to optically thick regime \citep{2018A&A...619A..61L}. This shift results in decreased X-ray flux and enhanced optical/UV emission. After early 2014, the boundary layer in T CrB became predominantly optically thick, likely representing the highest accretion rate ever recorded in the system \citep{2018A&A...619A..61L}.

X-ray emission in symbiotic binaries such as T CrB originates from the boundary layer between the WD and the AD \citep{2013A&A...559A...6L, 2018A&A...619A..61L, 2019MNRAS.489.2930Z}. Disk instabilities can cause large amounts of material to be dumped onto the WD over short timescales, resulting in an optically thick layer and marking the onset of enhanced activity \citep{2019MNRAS.489.2930Z}. The current high state is therefore attributed to a sustained increase in $\dot{M}_{\rm WD}$ \citep{2020ApJ...902L..14L, 2023A&A...680L..18Z}.

Throughout this period, the evolution of X-ray flux was strongly anti-correlated with the optical light curve, a relationship consistent with theoretical models of disk accretion in which higher accretion rates yield lower X-ray-to-optical flux ratios \citep{2024MNRAS.532.1421T}. Thus, the observed optical brightening is most plausibly driven by increased accretion onto the WD \citep{2018A&A...619A..61L, 2019ApJ...884....8L}. We agree with these interpretations.

\begin{figure*}\begin{center}
\includegraphics[width=17.0cm]{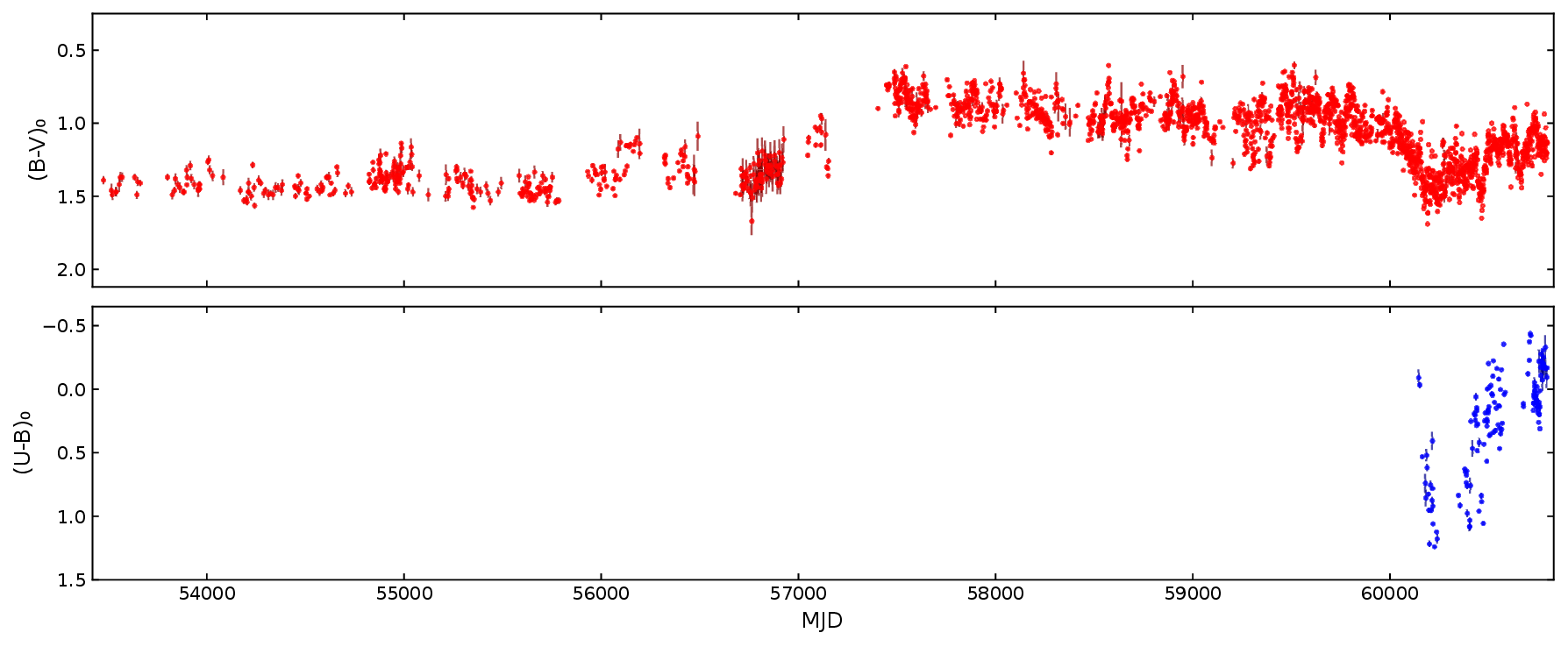}
\end{center}
\caption{Dereddened color evolution of T CrB from early 2005 to early 2025 based on AAVSO photometry. The top panel shows $(B-V)_0$ (red) and the bottom panel shows $(U-B)_0$ (blue). Colors were computed from near-simultaneous ($\pm$0.25 d) AAVSO $UBV$ measurements, adopting $E(B-V)=0.07$ \citep{2022NewA...9701859N}. For clarity, the dereddened color indices were averaged in 1-day bins.}
\label{fig:bvuv}
\end{figure*}

During the high state, several individual optical brightness peaks were observed and independently confirmed by \citet{2018A&A...619A..61L, 2023RNAAS...7..145M, 2024ATel16404....1M, 2025A&A...701A.176M}. In this study, however, we focus on the long-term average brightness trends. After accounting for the sinusoidal orbital modulation, the average $B$-band brightness increased by 0.9 mag compared to quiescence \citep{2025A&A...694A..85P}. The average brightness increases in the $V$, $R$, and $I$ bands were 0.38 mag, 0.22 mag, and 0.12 mag, respectively. As shown in Fig.~\ref{fig:bvuv}, the dereddened $(B-V)_0$ color evolution of T CrB indices remain relatively blue during the 2014–2023 high state, reflecting a significant contribution from hot accretion-driven emission. The average $(B-V)_0$ in quiescence is $1.41 \pm 0.04$, whereas during the high state it decreases to $0.93 \pm 0.02$.

\begin{figure*}\begin{center}
\includegraphics[height=15.4cm]{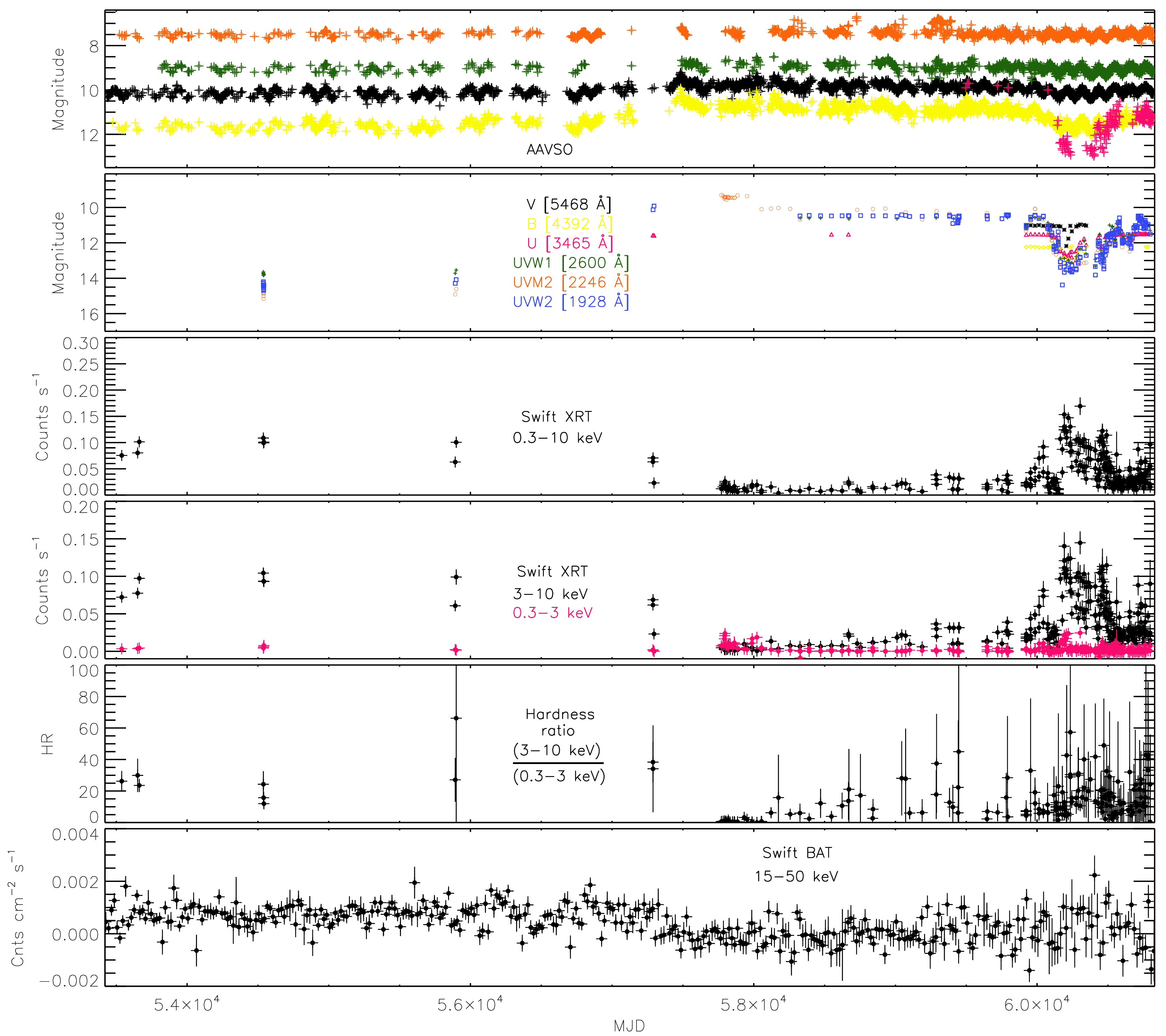}
\end{center}
\caption{From top to bottom: AAVSO light curves of T CrB from early 2005 to early 2025 in the following bands: I (green), R (orange), V (black), B (yellow), and U (red). Subsequent panels display: Swift/UVOT light curves in multiple filters; Swift/XRT 0.3 -- 10 keV light curve (PC mode); Swift/XRT light curves in 3 -- 10 keV (black) and 0.3 -- 3 keV (red); X-ray hardness ratio (3 -- 10 keV / 0.3 -- 3 keV); and the Swift/BAT light curve.}
\label{fig:sum band 01}
\end{figure*}

\begin{figure*}\begin{center}
\includegraphics[width=17.0cm]{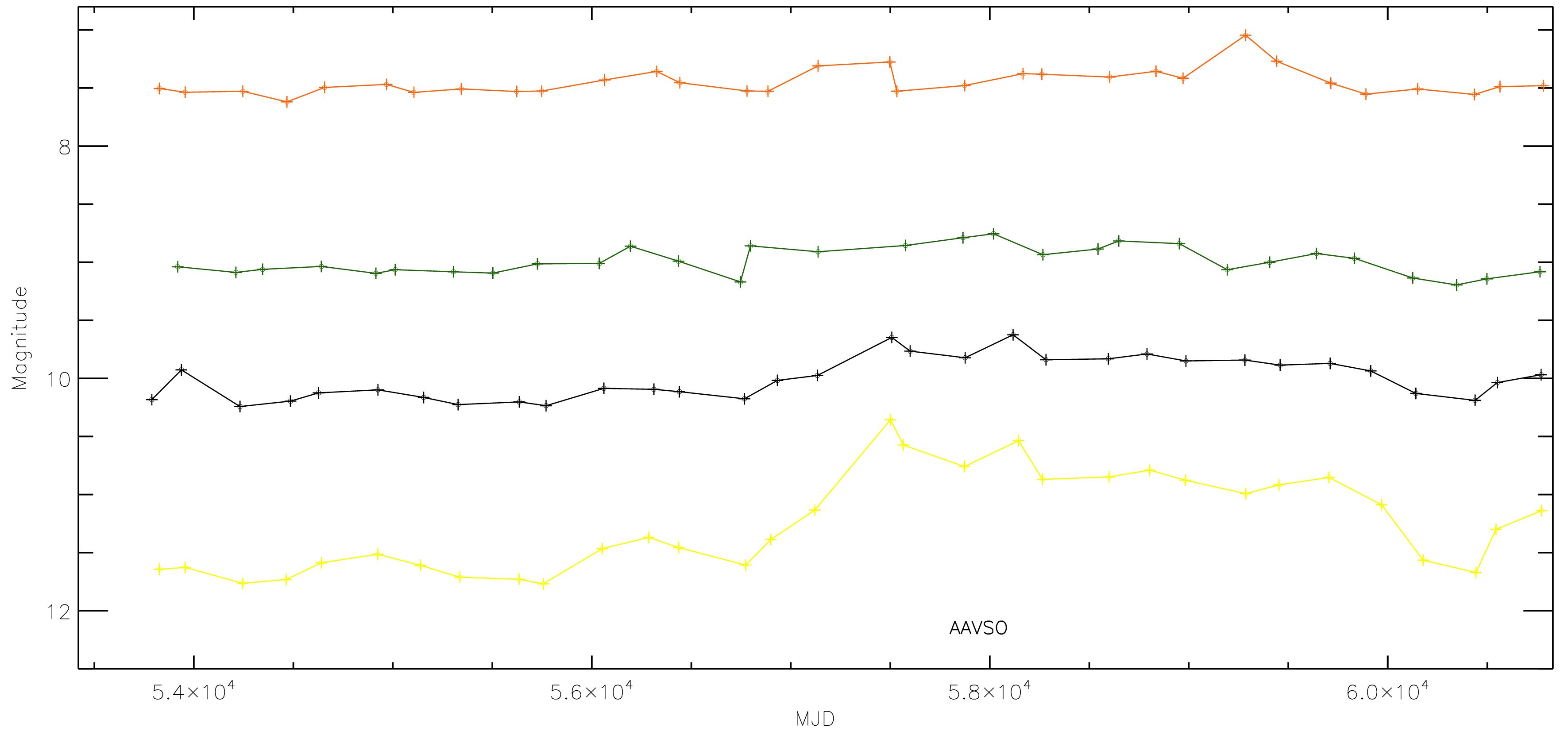}
\end{center}
\caption{AAVSO light curves of T CrB from 2005.78 to early 2025 in the I (green), R (orange), V (black), and B (yellow) bands. The data have been averaged over 227.58-day intervals, corresponding to the orbital period \citep{2025A&A...694A..85P}.}
\label{fig:irvu}
\end{figure*}

Soon after reaching the brightness maximum during the high state, T CrB begun a slow and steady descent in the $B$ band and $U$ band \citep{2025A&A...701A.176M}. As shown in Fig.~\ref{fig:irvu}, following the initial brightening in the $B$ band, the average brightness gradually declined throughout the remainder of the high state. This decline is anti-correlated with the X-ray emission in the 0.3–10 keV range (see Fig.~\ref{fig:sum band 01}).

X-ray count rates in the 0.3–10 keV band provide additional support for this inverse relationship. On 2005.46, the count rate was $0.092 \pm 0.003$; it increased slightly to $0.109 \pm 0.003$ on 2005.78 and remained high at $0.102 \pm 0.003$ on 2008.20. However, it declined to $0.071 \pm 0.004$ by 2011.92, $0.054 \pm 0.003$ by 2015.73, and reached a minimum of $0.012 \pm 0.001$ in 2017. This steady decline, apart from the slight early rise, spans more than a decade and coincides with the system’s optical brightening, reinforcing the observed anti-correlation between optical and X-ray flux.

In the UV bands, a similar trend is evident. Between 2008.20 and 2011.92, the average UVW1 (2600 \AA) magnitude increased by 0.13 mag, and the UVM2 (2246 \AA) magnitude by 0.09 mag, without accounting for orbital modulation. During the 2005–2013 quiescent phase, the B-band light curve of T CrB exhibits a slow but measurable increase in brightness at a rate of $\Delta B = -0.04$ mag yr$^{-1}$ \citep{2025A&A...701A.176M}. As shown in Fig.~\ref{fig:irvu}, the B-band brightness does not increase monotonically over this interval, but rather displays gradual fluctuations with alternating rises and declines. In contrast, the AAVSO light curves in the V, R, and I bands over the same period (2005–2013) do not exhibit a consistent upward trend (see Fig.~\ref{fig:irvu}). These results imply that UV and B-band flux is more variable and possibly anti-correlated with X-ray emission even during quiescence. Moreover, shorter wavelengths exhibit more pronounced variability, suggesting greater sensitivity to accretion-related changes.

From 2008.20 through the high state, the average brightness in the UVW1 band increased by 3.01 mag, and by 4.60 mag in the UVM2 band. Like the optical bands, UV brightness increased more strongly at shorter wavelengths, highlighting a consistent pattern across the spectrum during the transition from quiescence to the ctive state.

As shown in Fig.~\ref{fig:sum band 01}, the 0.3–3 keV X-ray count rates were relatively high and stable during 2017.05–2017.83 but remained low for the rest of the high state. Consequently, the hardness ratio (3–10 keV / 0.3–3 keV) primarily reflects variability in the 3–10 keV range and tracks closely with its evolution.

Since the onset of optical brightening, the X-ray spectrum has softened significantly. The hardness ratio (2–10 keV / 0.3–2 keV) declined by nearly two orders of magnitude by early 2018 \citep{2018A&A...619A..61L}. On 2018 January 30, the X-ray emission was harder than on 2017 February 23 \citep{2019MNRAS.489.2930Z}. During 2017.05–2017.83, the hardness ratio remained low (0.10–3.24), but gradually increased through 2021.64, and again between 2022.18 and 2022.60 (see Fig.~\ref{fig:sum band 01}).

The Swift/BAT hard X-ray (15–50 keV) light curve shows minimal flux during 2017.05–2017.83, followed by a gradual increase. This pattern closely resembles the evolution of the 3–10 keV X-ray light curve, further emphasizing the consistent multi-band behavior during the high state.

\subsubsection{Properties of the pre-eruption dip state} \label{sec:dip}
The 2023–2024 pre-eruption dip marks the most significant multi-wavelength transition observed in T CrB over the past eight decades. Its onset around MJD 60050 (April 2023) coincided with the decline from the prolonged 2014–2023 high state. The fading proceeded steadily over $\approx$ 130 days, reaching minimum brightness near MJD 60180 (August 2023).

During the pre-eruption dip in 2023–2024, the ellipsoidal variation in the $B$-band light curve became more clearly defined compared to the preceding high state. This modulation subsequently diminished after the dip phase.

Although the timing of the optical brightness minimum during the pre-eruption dip has been reported by \citet{2024ATel16404....1M, 2025A&A...701A.176M}, this study focuses on the average brightness across the dip. We compute the average magnitudes over a 113.79-day interval (MJD 60253.105–60366.895), corresponding to half the orbital period (227.58 days), centered on the bottom of the dip.

The average brightness decreased by the following amounts (after accounting for orbital modulation): 1.364 mag in the $U$ band, 0.738 mag in $B$, 0.293 mag in $V$, 0.148 mag in $R$, and 0.074 mag in $I$. This wavelength-dependent fading is consistent with a color-dependent decline, most prominent in the $B$ band and progressively smaller at longer wavelengths \citep{2025ApJ...989...78S}. A pronounced reddening develops during the 2023–2024 pre-eruption dip, with $(U-B)_0$ showing the largest change (see Fig.~\ref{fig:bvuv}), and an average $(B-V)_0$ of $1.33 \pm 0.02$ during the dip. In the UV, the average brightness declined by 2.533 mag in the UVM2 band and 2.146 mag in UVW1, again considering the sinusoidal modulation at half the orbital period.

Interestingly, the average $V$-band brightness during the dip remained $\sim$ 0.087 mag above the typical quiescent level. Similar small excesses above quiescence were also observed in the $I$ (0.046 mag), $R$ (0.072 mag), $B$ (0.162 mag), and UV bands.

As illustrated in Figs.~\ref{fig:sum band 01} and \ref{fig:sum band 02}, the 0.3–10 keV X-ray count rate generally increased during the fading phase of the pre-eruption dip (from its onset to the optical minimum), consistent with the findings of \citet{2024ATel16404....1M} that reported a steady rise in X-ray flux at the end of the high state. Subsequently, the count rate decreased during the post-dip recovery phase (from the optical minimum to the end of the dip).

The 0.3–10 keV X-ray light curve exhibits an "outburst-like" profile with three peaks during the pre-eruption dip. Notably, these peaks do not coincide with the maxima expected from sinusoidal modulation (i.e., phase = 0.25), suggesting intrinsic variability. The three peaks are present in both the 0.3–3 keV and 3–10 keV light curves, but the 3–10 keV band exhibits larger amplitude variations. Consequently, the hardness ratio increases and then decreases, producing a peak-like evolution during this interval. A similar peak-shaped trend is observed in the 15–50 keV hard X-ray light curve.

Throughout the pre-eruption dip, both the soft (0.3–10 keV) and hard (15–50 keV) X-ray light curves exhibit an anti-correlated relationship with the optical and UV brightness. Furthermore, the amplitude of the brightness decline across the optical and UV bands increases with decreasing wavelength, emphasizing the color-dependent nature of the dip. 

After the minimum of the 2023–2024 dip, both $(B-V)_0$ and $(U-B)_0$ gradually trend back toward bluer values, although neither returns to the high-state levels; a smaller but distinct reddening episode is again present. As illustrated in Fig.~\ref{fig:sum band 01} and Fig.~\ref{fig:sum band 02}, a second, lower-amplitude dip is visible in the UV bands between September 2024 and February 2025, hereafter referred to as the “second dip,” following the main pre-eruption dip. Although the sampling is sparser, both the UVW1 and UVM2 bands show declines of $\approx$0.8–1.1 mag relative to the preceding maximum, while the B band exhibits a modest $\sim$0.25 mag decrease. The event lasted for $\sim$150 days, comparable to roughly half an orbital cycle, despite a gap in Swift coverage during this interval. In contrast to the pronounced reddening that developed during the 2023–2024 pre-eruption dip, a smaller secondary reddening is also apparent during the 2024–2025 dip. This repetition indicates that the feature is intrinsic to the system rather than an artifact of orbital sinusoidal modulation.

\subsection{Timing analysis}
\label{sec:Timing}

We applied the Lomb-Scargle periodogram (LSP) method \citep{1982ApJ...263..835S} to search for periodicities in the X-ray light curves of T CrB. Specifically, we used the lomb package developed by Thomas Ruf\footnote{\url{https://cran.r-project.org/web/packages/lomb/index.html}}, which computes the LSP for unevenly sampled time series.

We separately analyzed the XRT light curves in the 0.2–0.6 keV, 0.3–3 keV, 3–10 keV, and 0.3–10 keV energy bands. No statistically significant periodicity was detected in any of these energy ranges. We did not expect to recover the previously reported 6000–6500 s periodic variability in the 0.2–0.6 keV band \citep{2019MNRAS.489.2930Z} using Swift/XRT, as these periods are too close to Swift’s orbital period of 5754 s. 

The Swift/BAT hard X-ray transient monitor has been detecting T CrB since 15 February 2005. We analyzed the BAT light curves in the 15–50 keV band spanning from 15 February 2005 (MJD 53416) to 20 July 2025 (MJD 60876), covering approximately 32.78 cycles of the 227.58-day orbital period and 65.56 cycles of the ellipsoidal modulation period. Neither the orbital period nor the ellipsoidal modulation period was detected in the orbital- or daily-binned light curves.

%

%
\section{Discussion}
\label{sec:discussion}
Our analysis builds upon earlier works \citep{2018A&A...619A..61L, 2020ApJ...902L..14L, 2019MNRAS.489.2930Z} by extending the Swift and AAVSO monitoring of T CrB through 2025. This 20-year dataset captures the full evolution of the system, from quiescence to the 2014–2023 high state and into the 2023–2025 pre-eruption dip sequence. The continuous multi-band coverage allows a quantitative characterization of the anti-correlation between optical/UV and X-ray fluxes and of the wavelength-dependent amplitudes of both brightening and fading. In contrast to previous studies that focused primarily on the 2014–2015 high state, we identify a new, secondary dip in late 2024 and demonstrate that the anti-correlation persists even during this later stage. Additionally, we present the first long-term dereddened color evolution of T CrB over two decades. These results provide a unified picture in which both the high and dip states are manifestations of variable mass-accretion rates and changing optical depth of the boundary layer.

The increase in optical brightness from quiescence to the high state is likely driven by an enhanced accretion rate \citep{2018A&A...619A..61L, 2019ApJ...884....8L}. Similarly, the corresponding increase in UV brightness from quiescence to the high state may also be attributed to enhanced accretion onto the WD.

\citet{2023MNRAS.524.3146S} suggested that a pre-eruption dip similar to the one observed before the 1946 outburst could serve as an indicator of an impending nova, with the eruption expected about a year after the dip onset. In March–April 2023, T CrB began fading, which was interpreted as a potential pre-eruption dip \citep{2023ATel16107....1S, 2023BAAVC.196....8T}, and based on this, an eruption was predicted for around 2024.4 $\pm$ 0.3. However, outburst has not occurred yet. \citet{2025MNRAS.541L..14M} found that the decline in the $B$-band during the 2023–2024 pre-eruption dip closely resembled that seen before the 1946 eruption, while the behavior in the $V$-band differed. During this recent dip, no deep decline occurred in the $V$-band; instead, the system returned to a typical quiescent level before re-brightening \citep{2025MNRAS.541L..14M}. Our analysis shows that the average brightness during the 2023–2024 dip remains higher than the historical quiescent level in both optical ($I$, $R$, $V$, $B$) and UV bands.

The 1946 pre-eruption dip, observed about 1 year before the outburst, featured a dramatic decline in the $V$ band—over 1.5 mag below the expected brightness of the red giant itself—suggesting an external obscuration mechanism. \citet{2023MNRAS.524.3146S} proposed circumstellar dust as the most plausible explanation. However, this hypothesis is challenged by the lack of stronger dimming at shorter (bluer) wavelengths, contrary to standard extinction laws. Alternative explanations have been considered. \citet{2023arXiv230804104Z} proposed the formation of a dense envelope ($\sim$ 1000–2000 km in radius) around the WD prior to the 1946 eruption. While this model could account for brightness declines, it fails to explain the magnitude of the dimming in the $V$ band during the 2023–2024 pre-eruption dip state. Additionally, it cannot account for the anti-correlated rise in X-rays seen during the 2023–2024 event.

\citet{2024ATel16404....1M} argued that the 2023–2024 dip was due to intrinsic variations in the brightness of the accretion disk rather than obscuration. They noted that the red giant’s contribution dominated the system’s light during this phase, and the accretion disk contributed negligibly. Our findings—showing clearer ellipsoidal modulation in the $B$ band and above-quiescence brightness levels—support this interpretation, implying that external dust obscuration may not be necessary to explain the dip. The $B$-band emission is primarily produced by the accretion disk and irradiation of the red giant by the WD. Therefore, a decrease in accretion rate leads to a corresponding decline in energy output from the disk \citep{2025BlgAJ..42...29S}.

The first X-ray detection of T CrB was made with the Einstein observatory \citep{1981ApJ...245..609C}, with subsequent studies conducted using multiple X-ray missions (Swift, RXTE, XMM-Newton, NuSTAR, and Suzaku) over the past two decades. T CrB exhibits hard and heavily absorbed X-ray emission \citep{2008ASPC..401..342L, 2009ApJ...701.1992K, 2016MNRAS.462.2695I, 2018A&A...619A..61L, 2019MNRAS.489.2930Z}, making it a typical member of the $\delta$-type X-ray symbiotic stars. In these systems, X-rays are thought to originate from the boundary layer between the accretion disk and the WD \citep{2013A&A...559A...6L, 2018A&A...619A..61L}. T CrB also shows strong X‑ray flickering (minute‑scale stochastic variability) in both soft ($< 10$) and hard ($> 10$ keV) X-rays \citep{2008ASPC..401..342L, 2016MNRAS.462.2695I}, consistent with accretion-driven processes. Boundary layer plasma temperatures showed dramatic changes from the quiescent state to the high state, that attributes to variations in mass accretion rate \citep{2024MNRAS.532.1421T}.

The observed anti-correlation between optical/UV and X-ray brightness, combined with a wavelength-dependent decline in brightness, parallels the behavior seen during the high state. Therefore, the pre-eruption dip may share the same underlying mechanism: variability in the mass accretion rate (see Sect. \ref{sec:high}). In this framework, the fading phase corresponds to a decline in $\dot{M}$, leading to lower optical/UV brightness and a transition to an optically thinner boundary layer—resulting in increased X-ray emission. This transition explains the observed anti-correlation between optical/UV and X-ray brightness throughout the dip. Conversely, the recovery phase marks a rise in $\dot{M}$, brightening the optical/UV bands while X-rays decline as the boundary layer becomes optically thick again. 

The presence of three peaks in the 0.3–10 keV X-ray light curve during the pre-eruption dip cannot be explained by orbital modulation or simple flickering. Instead, they are likely manifestations of variable mass accretion onto the WD, possibly triggered by disk instabilities or magnetic channeling.

During the high state, the structure of the boundary layer changed dramatically, consistent with a disk instability event responsible for the observed optical brightening \citep{2018A&A...619A..61L}. If the pre-eruption dip arises from the same physical mechanism as the high state, then the boundary layer may have undergone a similar or even more pronounced transformation. The duration of the pre-eruption dip ($\sim$ 1.5 years), comprising both the fading and recovery phases, is comparable to the $\sim$ 2-year optical brightening phase of the high state, further supporting this analogy.

The 2023–2024 pre-eruption dip may represent a pivotal stage in T CrB’s long-term evolution toward eruption. The depth of the dip increases systematically toward shorter wavelength indicates that the fading originates from the hot accretion component rather than from circumstellar extinction. Simultaneously, both the 0.3–10 keV and 15–50 keV X-ray fluxes rose sharply, producing a pronounced anti-correlation with the optical/UV light curves. Such behavior is consistent with a reduction in the accretion rate leading to a transition of the boundary layer from optically thick to optically thin emission.

Furthermore, T CrB's evolution in the X-ray hardness–intensity diagram (HID) resembles that of black hole binaries, accreting neutron stars, and active galactic nuclei (AGN), following hard state $\rightarrow$ hard-intermediate state (HIMS) $\rightarrow$ soft-intermediate state (SIMS) $\rightarrow$ soft state $\rightarrow$ SIMS $\rightarrow$ HIMS $\rightarrow$ hard state \citep{2024MNRAS.532.1421T}. However, during the 2023–2024 dip, T CrB deviated from this pattern, suggesting that the nova eruption may be imminent.

The second dip from September 2024 to February 2025 (following the primary pre-eruption dip) cannot be fully attributed to orbital modulation, it may share a similar formation mechanism–likely tied to variations in accretion rate. The correlated UV–optical behavior, the istinct reddening of $(B-V)_0$, and the simultaneous increase in soft-X-ray flux suggest that this second dip is not due to geometric obscuration but rather reflects a renewed, short-term reduction of the accretion rate and a temporary return to an optically thinner boundary layer. Its occurrence shortly after the main 2023–2024 dip indicates that the system may be undergoing a sequence of thermal–viscous instabilities in the inner disk as the white dwarf envelope approaches critical ignition conditions.

Therefore, the 2023–2025 dips is not merely a photometric fluctuation but a critical diagnostic of pre-nova accretion physics. Its detailed characterization provides the first modern observational counterpart to the fading recorded before the 1946 eruption. Taken together, these properties indicate that the pre-eruption dip represents a critical readjustment in the mass-transfer process as the system approaches thermonuclear runaway–analogous to, but more complex than, the fading recorded prior to the 1946 eruption. The dips therefore provide a valuable empirical probe of the conditions immediately preceding a nova outburst in a symbiotic recurrent system.

The sustained enhancement of the accretion rate to $\dot{M} \approx (2\text{-}6) \times 10^{-8} M_{\odot} \text{ yr}^{-1}$ since 2014 implies that over the last decade the white dwarf has accumulated $\gtrsim (1\text{-}2) \times 10^{-7} M_{\odot}$ of fresh material. For a near-Chandrasekhar-mass (1.37~$M_\odot$) white dwarf, this envelope mass is sufficient to trigger thermonuclear runaway within a few years \citep{2001ApJ...558..323H, 2013ApJ...777..136W, 2015MNRAS.446.1924H, 2020NatAs...4..886H}. The observed pre-eruption dips can therefore be interpreted as phases of temporarily reduced mass transfer as the disk and boundary layer re-adjust to the accumulating pressure in the WD envelope. The corresponding increase in X-ray emission during the dips marks the transition to a thinner, hotter boundary layer, consistent with the behavior expected immediately before runaway ignition. Thus, the photometric and spectroscopic evolution of T CrB in 2023–2025 provides empirical evidence that the system is entering the final pre-nova stage. Given the significant structural and photometric changes observed during the 2023–2024 pre-eruption dip, including changes in the accretion rate and boundary layer, and the consensus in the literature predicting an imminent nova eruption, the likelihood that this dip is unrelated to the upcoming outburst is low.

\section{Summary and Conclusions}
\label{sec:conclusions}
We present a comprehensive multi-wavelength study of the symbiotic RN T CrB, utilizing Swift (XRT/BAT/UVOT) and AAVSO observations spanning 2005–2025. This time frame covers the system’s quiescent phase, the high state, and the recent pre-eruption dip. The principal findings and conclusions of our work are as follows.

Analysis of the optical light curves in the I, R, V, and B bands confirms that the amplitude of brightening during the initial optical brightening of the high state is wavelength-dependent: the shorter the wavelength, the greater the flux increase relative to quiescence. A similar trend is evident in the UV bands, where shorter wavelengths exhibit more significant variability than longer wavelengths, both within the UV domain and when compared to the optical. We also find evidence suggesting that the UV emission is anti-correlated with the X-ray flux even during the quiescent state. During the high state, X-ray fluxes in both the 0.3–10 keV and 15–50 keV bands rose gradually after reaching a minimum in 2017. The B-band light curve remained anti-correlated with the X-rays beyond the initial optical brightening, supporting a scenario driven by variations in the accretion rate.

We find that the average optical and UV brightness levels during the bottom of the pre-eruption dip remained above those of the quiescent state. Throughout the entire pre-eruption dip, the brightness change in the optical and UV bands was more pronounced at shorter wavelengths. Furthermore, X-rays in both soft (0.3–10 keV) and hard (15–50 keV) bands were anti-correlated with the optical and UV emission during the entire pre-eruption dip. The presence of three peaks in the soft X-ray light curve during the dip likely reflects short-term fluctuations in the accretion rate. Following the main pre-eruption dip, a second, lower-amplitude dip was observed in the UV band between September 2024 and February 2025. Notably, no such secondary dip was observed after the pre-eruption dip of the 1946 outburst.

The pre-eruption dip (2023–2024) may have resulted from a decrease in the accretion rate, accompanied by a substantial transformation of the boundary layer. The observed variability across all bands suggests a transition from an optically thick to an optically thin boundary layer—mirroring the mechanism thought to drive the earlier high brightening phase. These results imply that the dip was not an isolated phenomenon, potentially heralding an imminent nova eruption.

The absence of modulation at either the orbital or ellipsoidal periods in the 15–50 keV hard X-ray light curves supports the interpretation that the hard X-ray emission arises from the compact boundary layer between the WD and the inner accretion disc, rather than from extended structures associated with the red giant or the wider binary geometry.

The 2023–2024 dip exhibits notable similarities with the pre-eruption behavior observed prior to the 1946 nova outburst. The subsequent recovery in brightness likely reflects resumed mass accumulation. Continued monitoring in the UV and X-ray bands, particularly during any forthcoming supersoft X-ray phase, will be essential to test the accretion-variation scenario.

This study presents the continuous 20-year, multi-wavelength record of T CrB encompassing quiescent, high, and pre-eruption dip phases. The analysis reveals two consecutive dips (2023–2024 and 2024–2025) characterized by wavelength-dependent fading and anti-correlated X-ray brightening, both consistent with temporary reductions in the accretion rate. The persistence of these transitions, together with the inferred accumulated envelope mass on the near-Chandrasekhar white dwarf, supports the interpretation that T CrB is approaching a thermonuclear runaway. These results establish a coherent accretion-variation framework linking the high and dip states and provide the comprehensive pre-eruption view yet obtained of a symbiotic recurrent nova on the verge of outburst.

Overall, T CrB’s sequence of brightening and fading episodes highlights the key role of accretion variations in modulating emission across the optical, UV, and X-ray regimes. The system provides a rare opportunity to investigate pre-eruption dynamics in long-period WD binaries. Future multi-wavelength observations, spanning from $\gamma$-rays to radio, during the anticipated nova eruption will offer a unique means to constrain the evolution of the accretion rate and to further advance our understanding of RN physics.

\section{Data availability}
\label{sec:data}
The Swift data presented in this article were obtained from the Mikulski Archive for Space Telescopes (MAST) at the Space Telescope Science Institute. The specific observations analyzed can be accessed via [doi: 10.17909/h2jm-m826]{https://doi.org/10.17909/h2jm-m826}.

\section*{Acknowledgements}
We extend our gratitude to observers worldwide for their valuable contributions to the AAVSO database. Additionally, this research utilized data provided by the UK Swift Science Data Centre at the University of Leicester. This work is supported by the High-level Talents Research Start-up Fund Project of Liupanshui Normal University (Grant No. LPSSYKYJJ202208), the Science and Technology Foundation of Guizhou Province (Grant No. QKHJC-ZK[2023]442), the Discipline-Team of Liupanshui Normal University (Grant No. LPSSY2023XKTD11), the Liupanshui Science and Technology Development Project (Grant No. 52020-2024-PT-01), the National Natural Science Foundation of China (Grant No. 12303054, 12393853, 12003020), the National Key Research and Development Program of China (Grant No. 2024YFA1611603), the Yunnan Fundamental Research Projects (Grant Nos. 202401AU070063, 202501AS070078), and the International Centre of Supernovae, Yunnan Key Laboratory (No. 202302AN360001), the Shaanxi Fundamental Science Research Project for Mathematics and Physics (Grant No. 23JSY015). RZS acknowledges the support from the China Postdoctoral Science Foundation (Grant No. 2024M752979).

\bibliographystyle{aa}
\bibliography{TCrB.bib}

\end{document}